\documentclass{article}

\usepackage{arxiv}

\usepackage{authblk }
\usepackage[utf8]{inputenc} 
\usepackage[T1]{fontenc}    
\PassOptionsToPackage{hyphens}{url}
\usepackage[colorlinks=true]{hyperref}      
\usepackage{url}            
\usepackage{booktabs}       
\usepackage{amsfonts}       
\usepackage{nicefrac}       
\usepackage{microtype}      
\usepackage{lipsum}
\usepackage[sort&compress, numbers]{natbib}
\usepackage{booktabs,multirow,multicol}
\usepackage{amsmath,amsthm}
\usepackage{algorithm,algpseudocode}
\usepackage{subcaption}
\usepackage{rotating} 

\usepackage[dvipsnames]{xcolor}
\usepackage{todonotes}

\hypersetup{colorlinks,breaklinks,
            citecolor=[rgb]{.05,.05,1.0},
            urlcolor=[rgb]{0,0.5,0.5},
            linkcolor=[rgb]{0,0.5,0.5}}

\title{Weekly sequential Bayesian updating improves prediction of deaths at an early epidemic stage}

\author[1,2,3,4]{Pedro Henrique da Costa Avelar}
\author[1]{Natalia del Coco}
\author[2]{Luis C. Lamb}
\author[3]{Sophia Tsoka}
\author[1,3,5]{\\Jonathan Cardoso-Silva \thanks{\bf Corresponding author. E-mails:  J.Cardoso-Silva@lse.ac.uk}\hspace{0.5em}}

\affil[1]{Data Science Brigade, Porto Alegre, Brazil \thanks{Part of this work was done while previously affiliated with this organisation. None of the authors hold this affiliation any longer.}}
\affil[2]{Institute of Informatics, Federal University of Rio Grande do Sul, Porto Alegre, Brazil}
\affil[3]{Department of Informatics, King's College London, London, United Kingdom}
\affil[4]{Machine Intellection Department, Institute for Infocomm Research, A*STAR, Singapore}
\affil[5]{Data Science Institute, London School of Economics and Political Science, London, United Kingdom}

\begin{document}
\flushbottom
\maketitle

\begin{abstract}
\textbf{Background:} Following the outbreak of the coronavirus epidemic in early 2020, municipalities, regional governments and policymakers worldwide had to plan their Non-Pharmaceutical Interventions (NPIs) amidst a scenario of great uncertainty. 
At this early stage of an epidemic, where no vaccine or medical treatment is in sight, algorithmic prediction can become a powerful tool to inform local policymaking.
However, when we replicated one prominent epidemiological model to inform health authorities in a region in the south of Brazil, we found that this model relied too heavily on manually predetermined covariates and was too reactive to changes in data trends.

\textbf{Methods:} 
Our four proposed variations of the original method allow accessing data of daily reported infections and take into account the under-reporting of cases more explicitly. 
Two of the proposed versions also attempt to model the delay in test reporting.
We simulated weekly forecasting of deaths from the period from 31/05/2020 until 31/01/2021.
First week data were used as a cold-start to the algorithm, after which weekly calibrations of the model converged in fewer iterations.
That workflow allowed us to run a lighter version of the model after the first calibration week. 
Google Mobility data, weekly updated, were used as covariates to the model at each simulated run.

\textbf{Findings:} 
The changes made the model significantly less reactive and more rapid in adapting to scenarios after a peak in deaths is observed. 
Assuming that reported cases were under-reported greatly benefited the model in its stability, and modelling retroactively-added data (due to the ``hot'' nature of the data used) had a negligible impact on performance.

\textbf{Interpretation:} Although not as reliable as death statistics, case statistics, when modelled in conjunction with an ``overestimate'' parameter,  provide a good alternative for improving the forecasting of models, especially in long-range predictions and after the peak of an infection wave.
\end{abstract}


\section*{Introduction}

The World Health Organization (WHO) declared COVID-19 a global pandemic in mid-March 2020, prompting countries to take actions to reduce the spread of the virus in view of the serious respiratory problems that require specialized care in Intensive Care Units (ICU) \cite{Cucinotta2020, WHO2020-pandemic-declaration}.
Little was known about this new strain of coronavirus that threatened to overwhelm health systems, had forced several countries to lockdown, and had already been rapidly spreading across Brazil \cite{Burki2020}. 
With no drug treatments or vaccines in sight at that time and in the face of a lack of national measures to prevent the spread of the disease \cite{TheLancetSoWhat2020, Baqui2020}, governors and mayors had to decide independently on the implementation of non-pharmacological measures (NPI) \cite{Garcia2020}. 
Amid this scenario and despite the harsh inherent challenges of epidemic modelling, mathematical models offered a timely approach to help understand the regional dynamics of contagion of the disease and to predict how this health crisis could unfold in the weeks and months that followed \cite{Jewell2020, Estrada2020, Roda2020, Bertozzi2020, Sanche2020, Kucharski2020, Russell2021, Goldsztejn2020, Li2021, Brauner2020}.

A prominent mathematical model was introduced by the MRC Center for Global Infectious Disease Analysis group at Imperial College London in March 2020 \cite{Ferguson2020, Flaxman2020a}, along with the source code and a technical description of the equations \cite{Flaxman2020}.
This model sought, above all, to estimate the impact and effectiveness of NPI measures taken by European countries at that moment in time. 
Nonetheless, the model produces other results of interest, such as an estimated number of people infected by SARS-CoV-2 and the variations in the reproduction number $(R_{t})$ up until the current date. 

In this paper, we present four variations of the Flaxman et al model (here called \texttt{base} model) to forecast deaths by COVID-19 and overcome limitations we observed after replicating the model on a weekly basis to the seven macro-regions that compose the state of Santa Catarina, a southern state in Brazil.
Work on this project started in March 2020 and, as we replicated the algorithm every week, we noticed that the original model did not predict major trend changes in mortality data.
Most noticeably, this happened in weeks when the number of new reported infections showed a rapid increase or decrease.
Since the \texttt{base} model did not have access to this data, it could not anticipate trend changes in infection patterns.
The covariates used in the model were also inadequate because they limited the effective reproductive number $R_t$ to change only at manually predetermined time points, where a NPI measure came into effect \cite{Kuhbandner2020}.
Despite these limitations, the model was used elsewhere to estimate the impact of NPI measures in two Brazilian states \cite{Candido2020}.

Other obstacles, not related to the algorithm itself but which we also consider, are the under-notification of infected cases \cite{prado2020analysis}, and the delays in test reporting \cite{Gostic2020}, all common problems at the early stage of the pandemic. 
The data suggests an average delay of 5 days from RT-PCR test collection until the result is available in the official database of Santa Catarina state.
Consequently, the last week of data is almost guaranteed to be incomplete and uninformative. 

Our proposed alterations to the original method aim at overcoming the limitations mentioned above, allowing this mathematical model to be used more effectively for forecasting.
Additional equations and algorithmic strategies allow the model to use reported cases to estimate deaths by COVID-19.

\section*{Results}\label{sec:results}

\begin{figure}[!hbt]
\centering
\includegraphics[width=\linewidth]{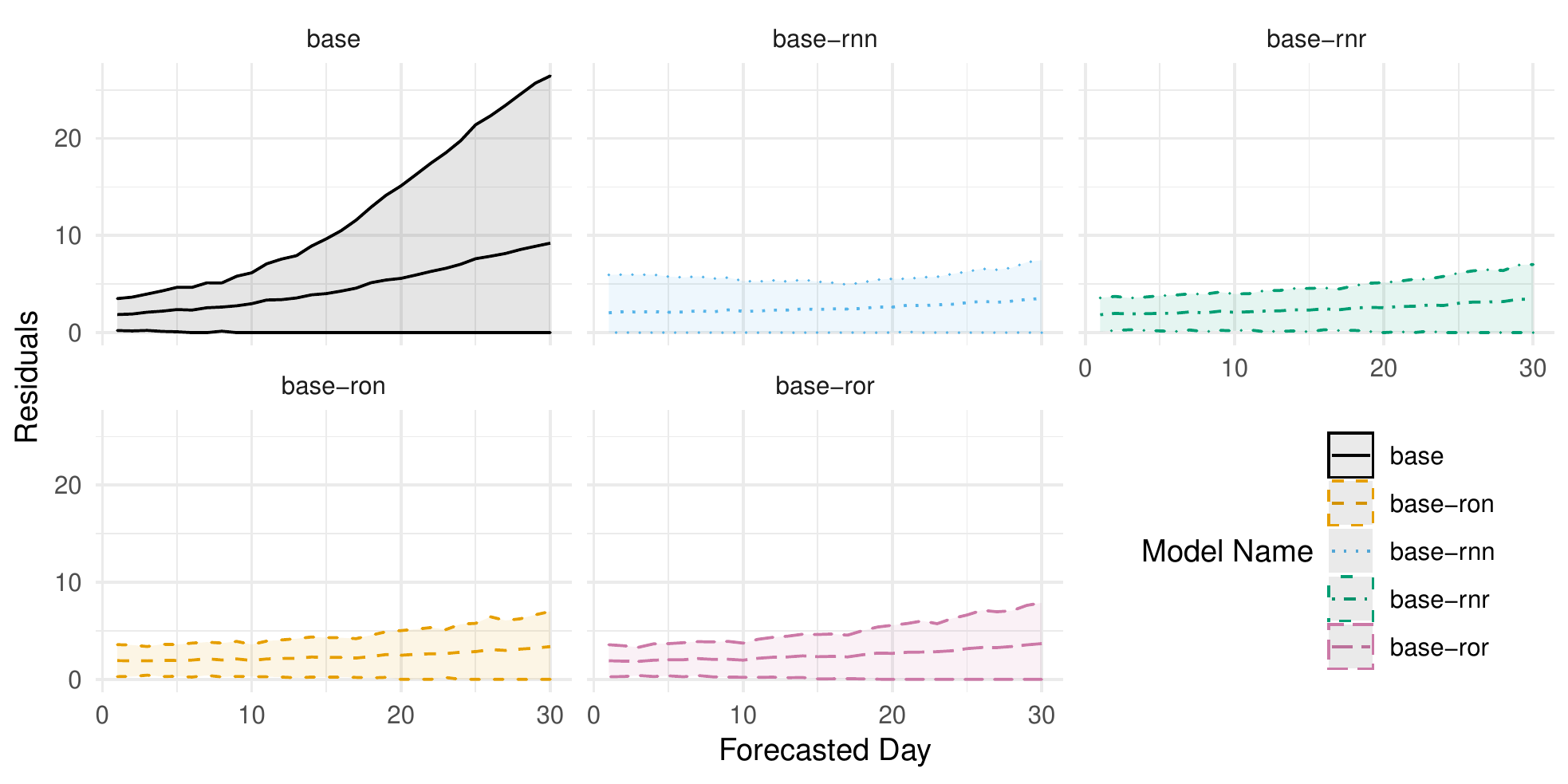}
\caption{Comparison of the proposed models to the \texttt{base} algorithm for forecasts over a time window of 30 days. The lines and areas show the average and standard deviation of absolute prediction error produced by each model and are here shown aggregated to the entire state of SC over the 35 weeks of modelled simulation.}
\label{fig:results-residuals}
\end{figure}

We simulated the \texttt{base} model and four variations of our proposed model, (\texttt{base-ron}, \texttt{base-rnn}, \texttt{base-ror}, \texttt{base-rnr}) described in Data and Methods section and summarised in Table \ref{tab:models}.
All models add the \textbf{newly reported infections} to the equations but they differ in whether we explicitly \textbf{overestimate}  infections in the equations and whether an estimated percentage of cases and deaths are added retroactively in the data before running the model to account for \textbf{delay in test reporting}. 

Each model produced a forecast for the 7 macro-regions of the state of Santa Catarina in weekly snapshots, starting from 31/05/2020 -- when daily snapshots of data became available in the official database system of the state -- until 31/01/2021.
At every simulated week, the models were fed with data exactly as it were available at that current date and produced a forecast of the number of deaths for the following 30 days.
The first week is used as a warm-up (to calibrate the method and allow us to run a lighter variant) and from the second week onward, the priors estimated in the previous week are provided to each corresponding model to help speed convergence.
More details about the proposed model, its variations, and our test workflow can be seen in the Data and Methods section.

\begin{figure}[!ht]
\centering
\includegraphics[width=\textwidth]{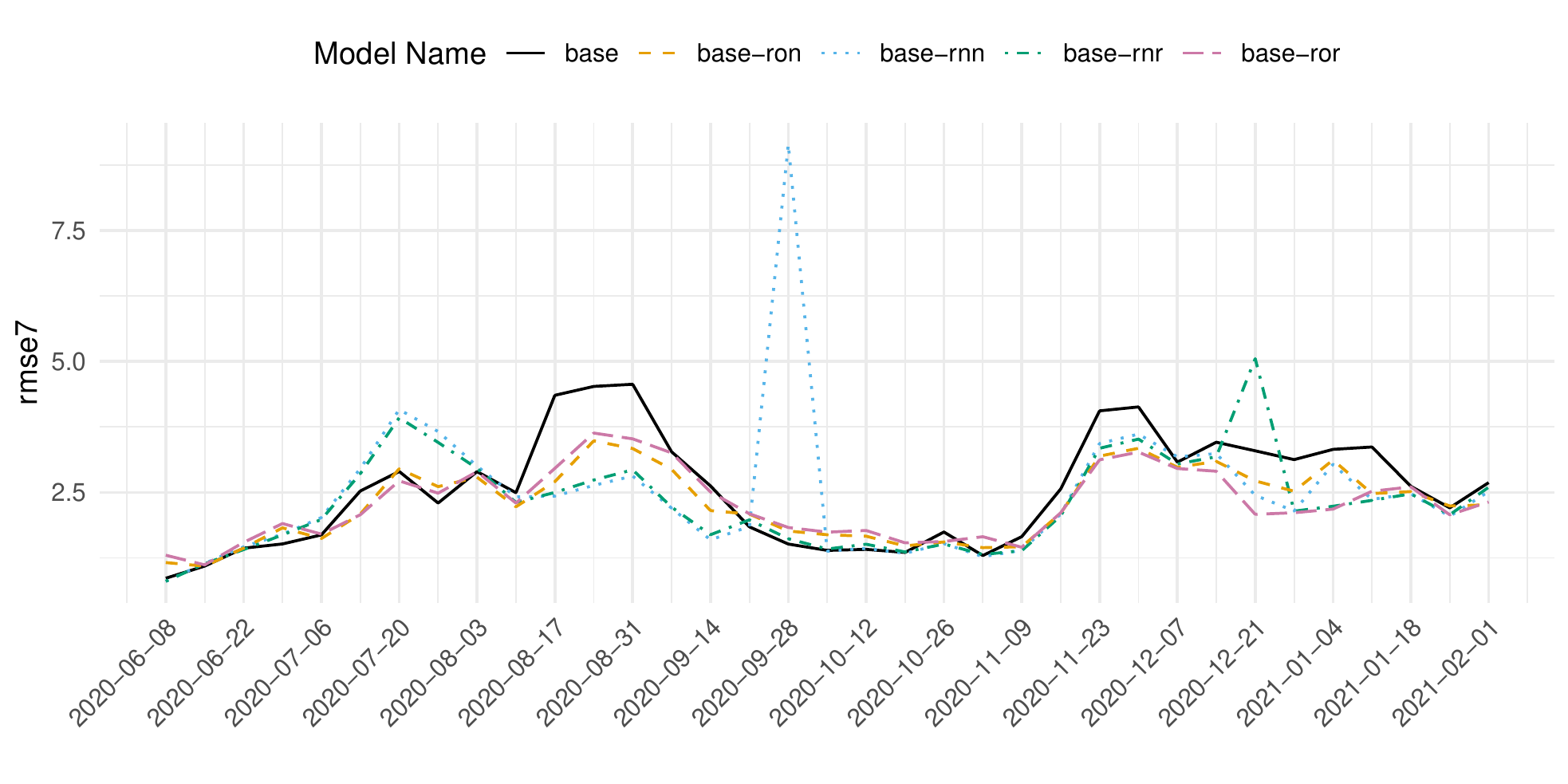}
\caption{Comparison of the modified models to the base algorithm for forecasts over a time window of 7 days.}
\label{fig:results-timeline-rmse7}
\end{figure}

\begin{figure}[!ht]
\centering
\includegraphics[width=\linewidth]{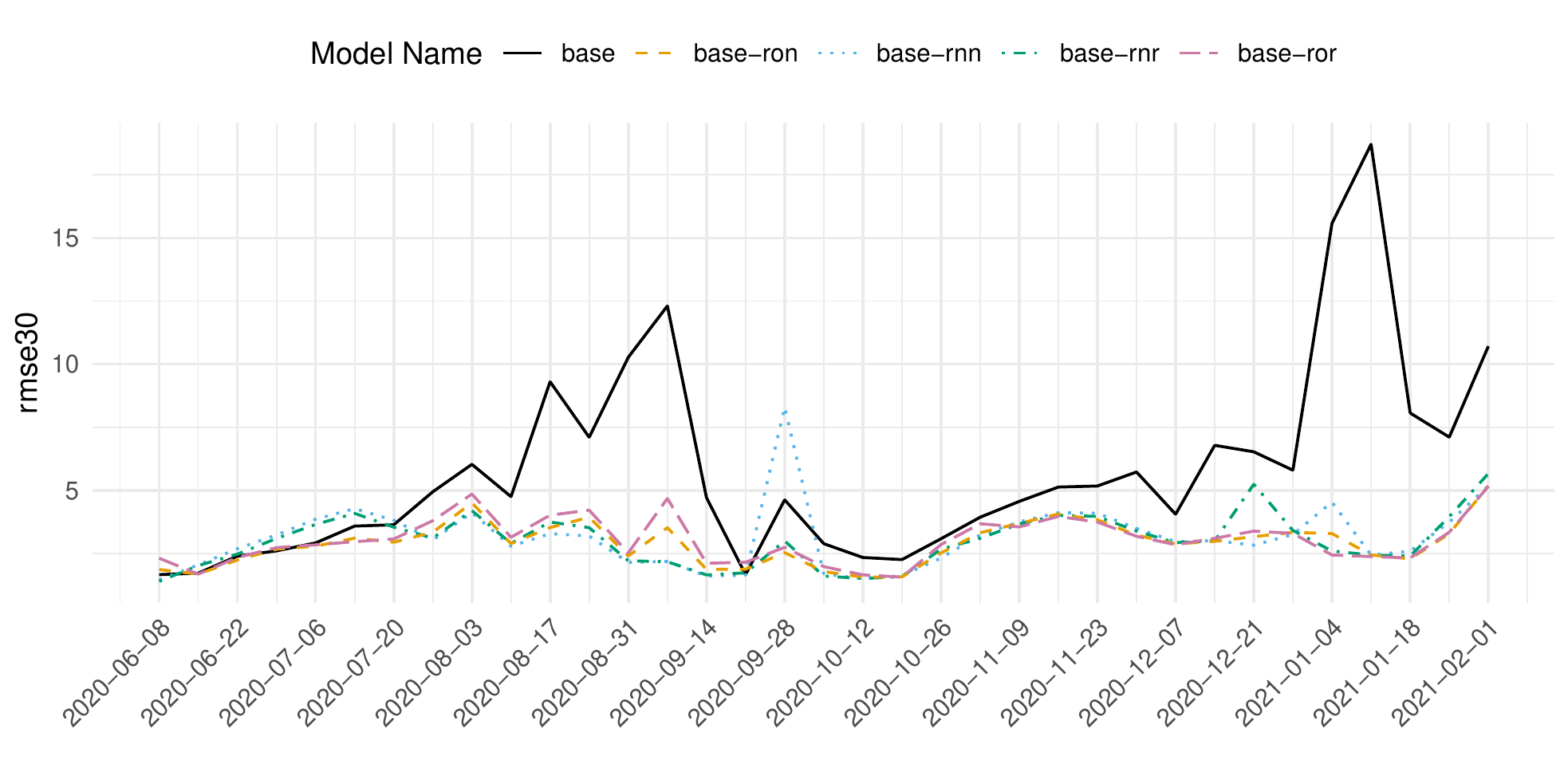}
\caption{Comparison of the modified models to the base algorithm for forecasts over a time window of 30 days.}
\label{fig:results-timeline-rmse30}
\end{figure}

On average, both original and alternative models had a similar prediction accuracy on the first 7-10 days of forecast but our proposed models outperformed the \texttt{base} model in the medium term. 
As shown in Figure~\ref{fig:results-residuals} which illustrates residual errors aggregated to the entire state of SC, the margin error of predictions made by \texttt{base} model grew wider over time whereas our models maintained a more stable error throughout the forecasting period.
Table \ref{tab:results} also provide a numerical comparison of these errors for a window of 7 and 30 days respectively and on Table \ref{tab:results-regions}, one could also inspect the results for each of the seven geographic regions that compose the state of Santa Catarina.

\begin{figure}[!ht]
    \centering
    \includegraphics[width=\linewidth]{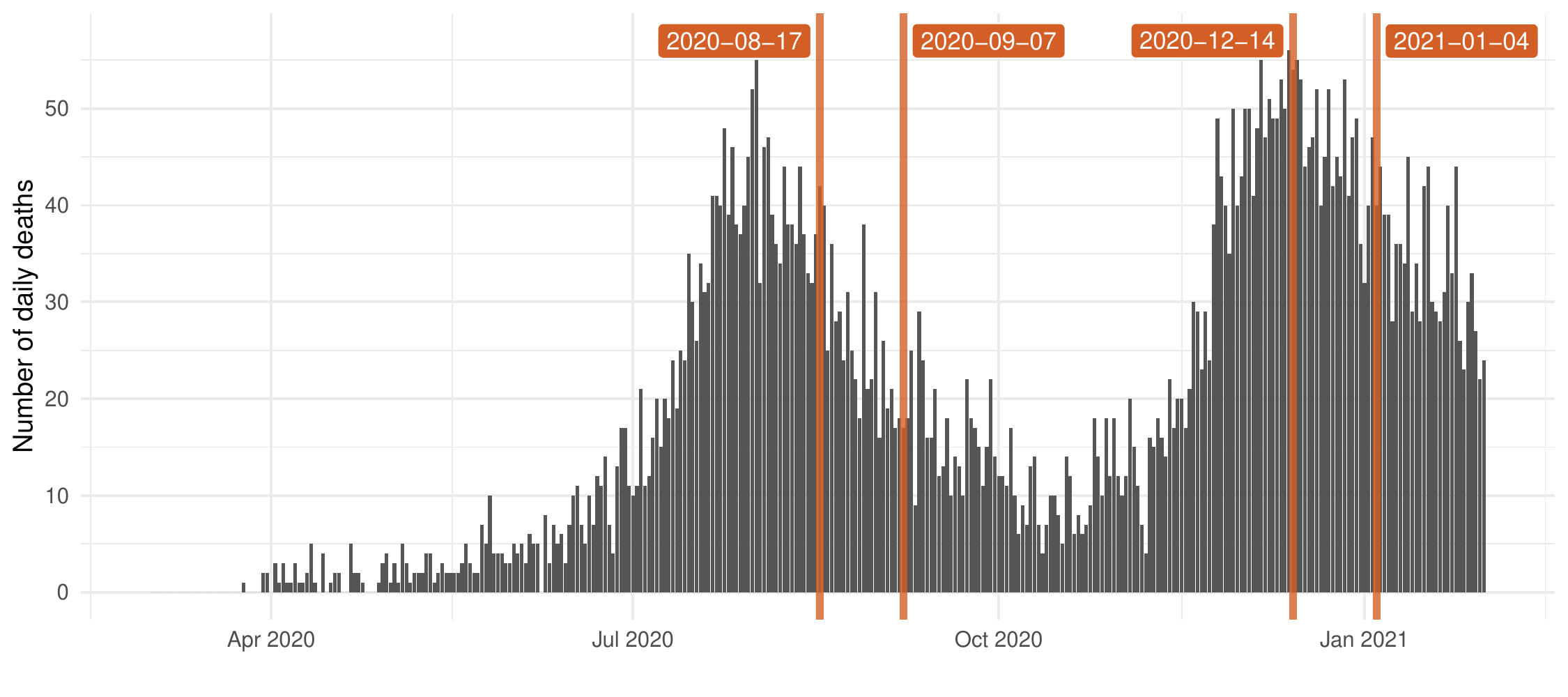}
    \caption{The curve of daily deaths by COVID-19 in the state of Santa Catarina. Highlighted are the dates in which prediction errors made by \texttt{base} model were higher.}
    \label{fig:deaths-sc}
\end{figure}

The gap between the baseline model and the proposed method over time is most noticeable at particular points in time, as indicated by RMSE plots for 7-day and 30-day forecasting windows of the models in Figures ~\ref{fig:results-timeline-rmse7} and ~\ref{fig:results-timeline-rmse30}, respectively. 
The \texttt{base} model showed the largest short-term error in the middle of August 2020 and at the end of November 2020.
On the 30-day window, the baseline algorithm is clearly making the worst predictions, particularly during August 2020 and the beginning of 2021 (Figure \ref{fig:results-timeline-rmse30}).
Compared to the curve of deaths in SC, shown in Figure \ref{fig:deaths-sc}, we observe that these higher errors were made on dates during or immediately after the peaks in the daily number of deaths (highlighted dates).
Diagnostic graphs produced by the model for these dates confirm that \texttt{base} was unable to reflect major changes in the trend of death data.
The model predicted that the number of infections was growing even though data regarding new reported infections already displayed a downward trend (Figures \ref{fig:base-model-abc-state-2020-08-10} - \ref{fig:base-model-abc-state-2020-09-07}).
One could contrast the diagnostic plots above to the ones obtained by \texttt{base-ron} model for the same dates in Figures \ref{fig:base-reported-model-abc-state-2020-08-10} - \ref{fig:base-reported-model-abc-state-2020-09-07}, where this misdirection in predicting death trend was not present in our proposed models.

\begin{figure}[!hbt]
    \centering
    \includegraphics[width=\linewidth]{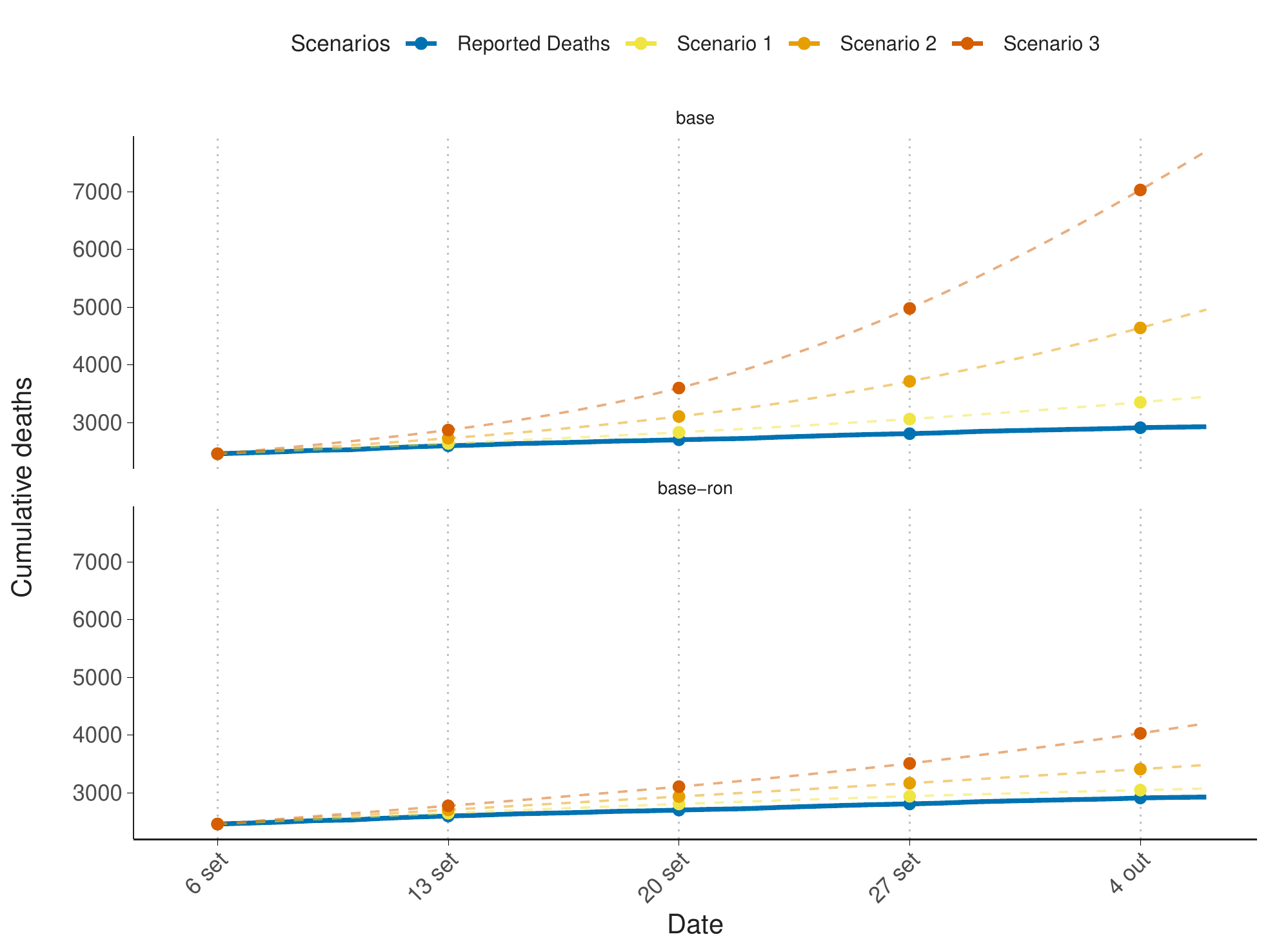}
    \caption{The forecast scenarios produced by \texttt{base} vs \texttt{base-ron} in 07/09 and the four following weeks. Scenarios 01 and 03 are the models' 95\% confidence interval around the average prediction indicated as Scenario 02. Our model clearly improved the predictions, providing a narrower confidence interval much closer to the real value.}
    \label{fig:followup09}
\end{figure}

Our methods outperform the baseline in nearly all runs, with the best algorithms being the ones with the "overestimate" variable: \texttt{base-ron} and \texttt{base-ror}.
These methods yielded the most stable predictions over both forecasting periods examined.
Interestingly, the posterior distribution of the ``overestimate'' parameter resulted in values much smaller compared to the priors we set ($\operatorname{overestimate}_m \sim \mathcal{N}(11.5, 2.0)$).
For example, the macro-region ``Foz do Rio Itajai'' had its mean close to 7.5, and the ``Grande Oeste'' macro-region had its mean lower than 2.5, as it can be seen in Figure \ref{fig:overestimate}.
The fact that the MCMC inference algorithm automatically converged to smaller values consistently across macro-regions suggests that, although present and significant, the sub-notification in the state of Santa Catarina was not as high as we had assumed.

On the other hand, \texttt{base-rnn} and \texttt{base-rnr} exhibited larger errors in predictions for certain periods in time, for example the middle of July 2020, later August 2020 or at the end of December 2020 (Figure \ref{fig:results-timeline-rmse7}).
Adding an estimate of the retroactive data did not seem to impact the predictive accuracy since this characteristic was present both in one of the best-performing models (\texttt{base-ror}) and in one of the less predictive ones, \texttt{base-rnr}.
Other assumptions embedded in each model variation can be found in the Materials and Methods Section and in Table \ref{tab:models}.

Another way to visualize these results is by comparing the graph of cumulative deaths with predictions made by \texttt{base} and one of the best performing models \texttt{base-ron} (Figure \ref{fig:followup09}).
Scenarios 01 and 03 show models' 95\% confidence interval around the average prediction indicated as Scenario 02. 
It is clear that our model has improved the predictions, providing a narrower confidence interval which was closer to the real value.

\section*{Discussion}

There is evidence in the literature that COVID-19 models are inefficient for long-range forecasting \cite{Moein2021inneficiency}. 
In this paper, however, we show that one can use ``hot'' data (i.e., the most recent data, that is constantly being updated and might still not be complete) to update a model on a weekly basis and achieve higher accuracy.
Such ``hot'' data come with own sets of issues, such as the unreliability, delays in updating, etc.
Knowledge of SARS-CoV-2 interactions in the human body, as well as social contagion dynamics of the disease, have not been elucidated entirely.
A lot is still unknown and epidemic modelling has to address real-time uncertainties and changes as closely as possible.
A model which does not take into account the fast pace production of data is bound to underperform when compared to a model based on up-to-date information. 
Our proposed models overcome some limitations of the original algorithm in Flaxman et al \cite{Flaxman2020} mostly by accessing data regarding daily reported infections and letting it account for under-reporting in a more explicit way.
We have also tested some variations in the algorithm to account for the delay within test collection and notification but these did not prove as useful in predicting the death curve in the state of Santa Catarina.

These alternative models, however, are not without their failings. 
While predictions have improved and some assumptions of the model could be confirmed by inspection of the data -- for example, the onset-to-death seems to follow a gamma distribution like the original model assumed -- there are just too many assumptions in the original model that have not been thoroughly validated\cite{Soltesz2020a} (for example, the $R_{0}$ parameter). 
Another issue is that, from the point of view of the optimisation algorithm, estimated $R_{t}$ values and estimated number of people daily infected are interchangeable.
The algorithm could reach two opposing configurations that are equally valid and optimal: one where the reproduction rate is low but there is a large pool of infected people in the population, and another separate solution in which the number of infected people is small but $R_{t}$ is larger.

Also, even though adding infection data into the model has given it more adaptability, the same data could make the model more fragile in the future.
If the infection dynamic changes (e.g. because of new more transmissible strains of the virus), the model might be biased to readjust its fitting of the historical data to compensate.  
In theory, this could be counteracted by more reliable epidemiological data from other sources (i.e., tracking the prevalence of new virus strains, information about age, or information on entry and exit into ICU), or by including even more granular mobility data, at the expense of the citizens' privacy, all of which are generally more expensive or infeasible to obtain.

New or data of higher resolution can alleviate some challenges in epidemic modelling but, importantly, new observations can help us revise assumptions of existing models in search of a more accurate description of real-world cases \cite{Moein2021inneficiency,JordanaCepelewicz2021hard}. 
Our proposed model is one step in that direction of scientific inquiry.
We show that a model can become more accurate by adding one more data source and a new assumption about under-reporting the tests, and therefore more useful for forecasting and decision making.
We intend to review the other assumptions built into the original model and continue to investigate how much the changes we have introduced are sustained in the face of new epidemic waves, new variants of the virus and new political measures that affect the dynamics of contagion.

\section*{Data and Methods}

\subsection*{Data}

Anonymized data on every confirmed case of COVID-19 in SC along with the date of onset of first symptoms, date of PCR test collection, date of death and municipality of residence were downloaded from the state government big data platform Plataforma BoaVista \cite{CIASC-SC2021}.
Estimated population in each of the seven regions of Santa Catarina were obtained from the Brazilian Institute of Geography and Statistics IBGE\cite{ibge2019pop} and mobility data for every city was downloaded from Google Mobility \cite{Aktay2020}.
The state of Santa Catarina has over 7 million inhabitants organised in 297 municipalities organised in 6 distinct geographical macro-regions distributed across 95 square kilometres of land area. 
The state government imposed suspensions of many economic activities after the first deaths were confirmed in SC in March 2020 but ended up relaxing social distancing measures, eventually leading to a decree in June 2020 after which municipal governments would be responsible for most decisions regarding NPI measures.
At the time of writing (24 March 2021), over 764,000 cases and over 9,800 deaths by COVID-19 had been confirmed, hospitals were fully occupied, and local news reported that at least 397 people were on the waiting list for ICU beds. \cite{G1-SC2021}.


\subsection*{Bayesian Model}
\label{sec:bayesian}

The baseline model (\texttt{base}) is almost identical to the Monte-Carlo (MCMC) model made available by Flaxman et al \cite{Flaxman2020}, with the exception of the covariates used.
Instead of manually curated non-pharmaceutical intervention measures, we used Google Mobility data as covariates to the model.
This strategy counteracts an implicit confirmation bias of the model  \cite{Kuhbandner2020} and have been used by the original group in a subsequent work \cite{Mellan2020}.
In contrast to European countries targeted by the original model, the State of Santa Catarina did not impose strict state-wide measures consistently during this period. 
Local governments (cities and regional associations of municipalities) were responsible for independently deciding on their social distancing measures \cite{Garcia2020}, which rendered changes in legislation impractical.
Since our Bayesian models rely on the same distributions as the \texttt{base} model, we will use the same symbols to mean the same distributions whenever possible. 
We refer the reader to their paper for a full description of the symbols \cite{Flaxman2020, Flaxman2020a}.
 
We now describe the equations and modifications of our proposed models, which are also summarised in Table \ref{tab:models}.
In mathematical terms, we observe daily deaths $D_{t,m}$ and daily cases $C_{t,m}$ for days $t \in \{1,\ldots,n\}$ and regions $m \in \{1,\ldots,M\}$. These values might be augmented in some models (e.g. \texttt{base-ror} and \texttt{base-rnr}) to a $D^{*}_{t,m}$ and $C^{*}_{t,m}$, which are explained later in this section. The objective function of the original model, and part of the objective of our models is given by:

\begin{equation}\label{eq:deaths-distribution}
    D^{*}_{t,m} \sim \operatorname{Negative Binomial}\left(d_{t,m}, d_{t,m} + \frac{d^{2}_{t,m}}{\Phi}\right),
\end{equation}

\noindent where $d_{t,m}$ represents the number of modelled cases, following:

\begin{equation}
    d_{t,m} = \operatorname{xfr}^{*}_{m} \sum_{\tau = 0}^{t-1}{c_{\tau,m} \pi_{t-\tau}},
\end{equation}

\noindent where $\operatorname{xfr}$ is either the region-specific case-fatality rate $\operatorname{cfr}$ (in models \texttt{base-rnn} and \texttt{base-rnr}) or the infection-fatality rate $\operatorname{ifr} \approx 0.0076$ estimated for Brazil in \cite{Mellan2020} (in models \texttt{base}, \texttt{base-ron}, and \texttt{base-ror}).
All models have an uncertainty factor added in the same way as in the original model.

A key component in all variations of the proposed algorithm is that, instead of relying solely on the death-based data to model the pandemic -- largely considered to be the most reliable source -- we also use the \textbf{reported infections} as a way to make the model less reactive.
In some model variations, we assume that the number of reported infections $C_{t,m}$ is under-reported and we model this by adding a normally distributed \textbf{overestimate} variable $\operatorname{overestimate}_m \sim \mathcal{N}(11.5, 2.0)$ to model under-reporting explicitly, following estimates that the number of COVID-19 cases in Brazil was about 11 times higher than what is officially reported \cite{prado2020analysis}.
So, instead of optimising only the original objective function present in Equation~\ref{eq:deaths-distribution}, we also rely on Equation~\ref{eq:cases-distribution} below to calibrate our model:

\begin{equation}\label{eq:cases-distribution}
    C^{*}_{t,m} \sim \operatorname{Negative Binomial}\left(p_{t,m}, p_{t,m} + \frac{p^{2}_{t,m}}{\Phi}\right), ~~~ p_{t,m} = c_{t,m} / \operatorname{overestimate}_{m}
\end{equation}

\noindent where $p_{t,m}$ estimates the actual number of people infected at time $t$ considering an overestimate.
$c_{t,m}$ represents the number of reported cases, and it depends on the same assumptions of susceptible individuals $S$, reproduction rate $R$, and generation distribution $g$ as the original model:

\begin{equation}\label{eq:cases-modelled}
    c_{t,m} = S_{t,m} R_{t,m} \sum_{\tau = 0}^{t-1}{ c_{\tau,m} g_{t-\tau}}
\end{equation}

\begin{sidewaystable}
    \centering
    \begin{tabular}{p{1.5cm}p{5cm}ccccc}
        \toprule
       & & \multicolumn{2}{c}{Optimisation Objectives} & \multicolumn{2}{c}{Model Interpretation} & \multicolumn{1}{c}{Model Inputs} \\
        Model Name & Description & $d_{t,m} \propto k_{d} * D^{*}_{t,m}$ & $c_{t,m} \propto k_{c} * C^{*}_{t,m}$ & $d_{t,m} \propto XFR * c_{t,m}$ & $c_{t,m}$ & $C^{*}_{t,m}, D^{*}_{t,m}$ \\
        \midrule
        $\operatorname{\texttt{base}}$
            & Baseline Model
            & $k_{d} = 1$
            & Not Used
            & $XFR = IFR$
            & Real Cases 
            & As reported \\[0.5em]
        
        $\operatorname{\texttt{base-ron}}$
            & Includes reported cases and overestimate infections
            & $k_{d} \approx 1$ $^{\star}$ 
            & $k_{c} \sim \mathcal{N}(11.5, 2.0)$ $^{\star\dagger}$ 
            & $XFR = IFR$ 
            & Real Cases $^{+}$ 
            & Augmented $^{\ddagger}$ \\
        
        $\operatorname{\texttt{base-rnn}}$
            & Includes reported cases but does not attempt to overestimate infections
            & $k_{d} \approx 1$ $^{\star}$
            & $k_{c} \approx 1$ $^{\star}$
            & $XFR = CFR$
            & Reported Cases
            & Augmented $^{\ddagger}$ \\
        
       $\operatorname{\texttt{base-ror}}$
            & Includes reported cases, model retroactive data and overestimate infections
            & $k_{d} \approx 1$ $^{\star}$
            & $k_{c} \sim \mathcal{N}(11.5, 2.0) ^{\star\dagger}$
            & $XFR = IFR$
            & Real Cases $^{+}$
            & As reported \\
        
        $\operatorname{\texttt{base-rnr}}$
            & Includes reported cases, model retroactive data but does not attempt to overestimate infections
            & $k_{c} \approx 1$ $^{\star}$
            & $k_{d} \approx 1$ $^{\star}$
            & $XFR = CFR$
            & Reported Cases
            & As reported \\
        \bottomrule
    \end{tabular}
    \caption{
    The models differ mainly in which objectives they optimise and how one can interpret the model. For our proposed models, we use infection data (with and without an estimate of under-reporting) to try to make the model less reactive, since data depending on deaths may only reflect the situation from a few weeks back. $d_{t,m}$ represents the number of predicted deaths and $D^{*}_{t,m}$ is the number of deaths used as an input to the model, in the same fashion $c_{t,m}$ and $C^{*}_{t,m}$ are the number of predicted cases and the number used as an input to the model.
    $^{\star}$ These values may not be exact, since the model has to take into consideration the number of both cases and deaths.
    $^{\dagger}$ $\mathcal{N}(11.5, 2.0)$ follows from estimates that the number of COVID-19 cases in Brazil was about 11 times higher than officially reported \cite{prado2020analysis}.
    $^{+}$ These cases are interpreted as the number of real cases as long as all the assumptions of the model hold true, which most likely they do not.
    $^{\ddagger}$ Cases $C_{t,m}$ and deaths $D_{t,m}$ from the last week of a data snapshot are augmented according to how historically these values had been retroactively changed, by having $C^{*}_{t,m} \approx C_{t,m} k_{c,t,m}$ and $D^{*}_{t,m} \approx D_{t,m} k_{d,t,m}$. }
    \label{tab:models}
\end{sidewaystable}

Our models have $K=7$ covariates. Six of them are the Google Mobility indicators, and the remaining covariate is the calculated percentage of the population of a region Susceptible to infection $S_{t,m}$. 
Since, in practice, Google Mobility data was unavailable for future dates, when simulating the weekly runs, we simply assumed that these covariates would remain constant from one week before the snapshot date.
The reproduction rate $R_{t,m}$ is assumed to vary with the covariates:

\begin{equation}\label{eq:Rt-modelled}
    R_{t,m} = R_{0,m} \exp^{-\sum_{k=1}^{K}{I_{k,t,m} (\alpha_{k,m} + \alpha_{k}^{*})} - S_{t,m} (\alpha_{pop,m} + \alpha_{pop}^{*})},
\end{equation}

\noindent the percentage of susceptible population $S_{t,m}$ (the seventh covariate) was modelled with a similar impact measure $\alpha_{pop}$ as the mobility data. 
On both the $\operatorname{\texttt{base}}$ and our proposed models we use an extra-region impact measure $\alpha_{k}$ as well as a per-region impact measure $\alpha_{k,m}$.
To consistently simulate the baseline algorithm, our simulations with $\operatorname{\texttt{base}}$ model did not include $S_{t,m}$ as a covariate, and we also used the same way of weighting these covariates, with both a state-wide $\alpha_{k}$ and a per-region $\alpha_{k,m}$.

\begin{figure}[!htbp]
    \centering
    \includegraphics[width=0.8\linewidth]{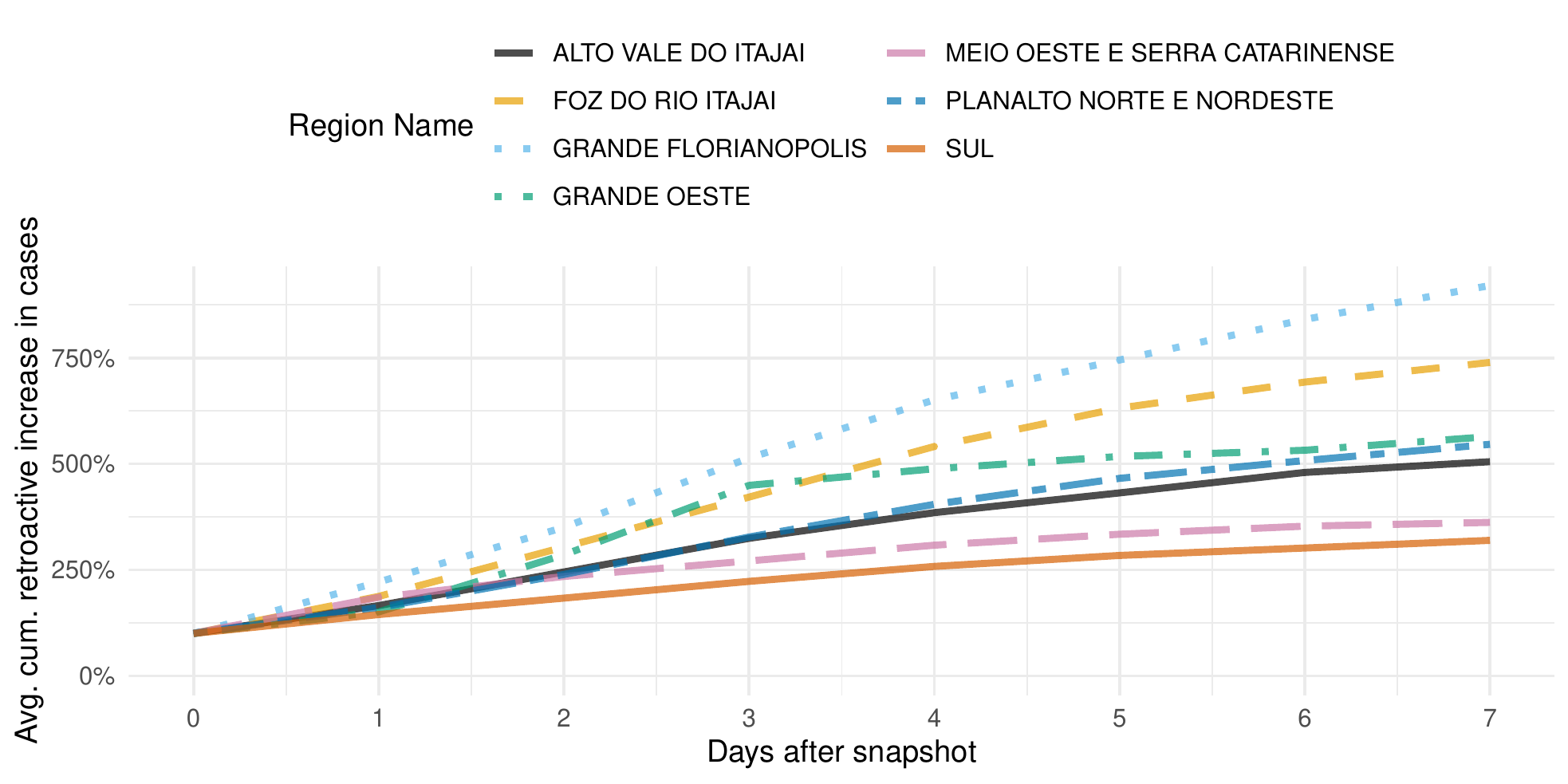}
    \caption{Distribution of cumulative retroactively-added cases on the last date.}
    \label{fig:cum-retroactive-cases}
\end{figure}

\begin{figure}[!htbp]
    \centering
    \includegraphics[width=0.8\linewidth]{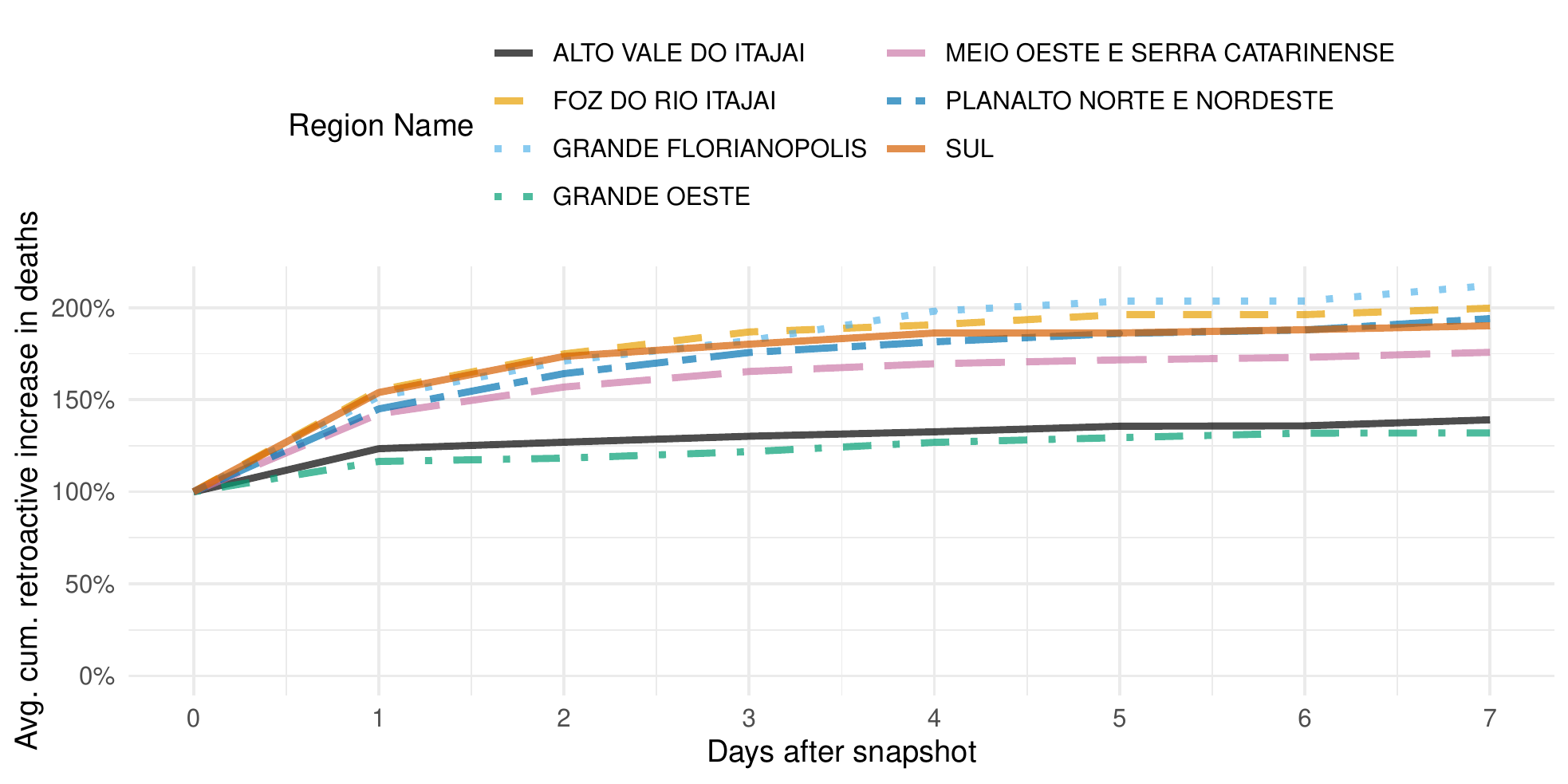}
    \caption{Distribution of cumulative retroactively-added deaths on the last date.}
    \label{fig:cum-retroactive-deaths}
\end{figure}

Some of our models (\texttt{base-base-ror} and \texttt{base-rnr}) also try to take into account the delay between PCR test collection and test result notification.
We have looked at how this reporting delay affects the data in the Santa Catarina state and results can be seen in Figure \ref{fig:cum-retroactive-cases} for cases and in Figure \ref{fig:cum-retroactive-deaths} for deaths, according to each region of the state. 
Notice from these plots that infections generally can be between 2.5x to 7.5x higher than their initial reported values, and thus from 60\% up to more than 85\% of cases are reported with 5 days of delay.
Deaths follow a similar, but less volatile, a pattern rarely passing values 2x higher than their initial reports, nonetheless possibly having 50\% of the data being reported after this period. 
While this confirms that deaths are the most reliable source of information, this also shows an issue in using such ``hot'' data for modelling: it can often be incomplete and lead to a false decrease in the number of cases and deaths.

In the aforementioned models, we increase the number of cases and deaths of the past week by a percent delta before running the model to try to compensate for this delay, the distribution of which can be seen in Figures~\ref{fig:retroactive-cases}~and~\ref{fig:retroactive-deaths}. This gives us:

\begin{equation}
  C^{*}_{t,m} = C_{t,m} (1 + \operatorname{\Delta{}retroactive}_{c,t,m})
\end{equation}

\noindent and
\begin{equation}
  D^{*}_{t,m} = D_{t,m} (1 + \operatorname{\Delta{}retroactive}_{d,t,m})
\end{equation}

Another minor notification was that the original model assumed the onset-to-death distribution to follow the distribution $\operatorname{Gamma}(17.8,0.45)$.
We kept the gamma function but we estimated the average and deviation from the data at each snapshot.
For reference, this value was close to $\operatorname{Gamma}(20.67,0.76)$ on the latest simulated weeks and the distribution of onset-to-death in Santa Catarina can also be seen in Figure \ref{fig:onset-to-death}.

\subsection*{Test Workflow and Validation}
\label{sec:test_workflow}

Our main goal here is to test which combination of inputs and equations would most accurately forecast the curve of deaths by COVID-19 for the seven demographic macro-regions within Santa Catarina. 
With this in mind, we planned the test to span the period 31/05/2020 to 31/01/2021, running the model as if it was run weekly using only the data available at that point in time. 
The results were then aggregated to compose the overall prediction for the entire state.

The models produce an average prediction which we compare to the ``ground-truth'' number of deaths using Root Mean Squared Error (RMSE) and the Mean Average Error (MAE) metrics (Results in Tables \ref{tab:results}-\ref{tab:results-regions} and Figures \ref{fig:results-timeline-rmse7} and \ref{fig:results-timeline-rmse30}).
Because of test notification delays discussed in the previous section and shown in Figures \ref{fig:retroactive-cases}-\ref{fig:retroactive-deaths}, we selected the data snapshot from 07/03/2021 to represent the ``ground-truth'' -- 1 month after the last run simulation.
We also wanted to estimate the error of the model given its confidence intervals and to measure this, we calculated RMSE and MAE of the upper and lower confidence intervals and took their average.  
These metrics RMSE\_conf and MAE\_conf provide a measure of how distant the borders of the confidence interval are from the truth values (shown in Tables \ref{tab:results}-\ref{tab:results-regions}).

\begin{algorithm}
    \begin{algorithmic}[1] 
        \Procedure{Test}{}
            \State $\operatorname{last-date} \gets \operatorname{NULL}$
            \For{$\operatorname{current-date} in \operatorname{dates}$}
                \If{$\operatorname{last-date} is \operatorname{NULL}$}
                    \State $\operatorname{run-and-save-model}(\operatorname{current-date}, \operatorname{initial-hyperparameters})$
                \Else
                    \State $\operatorname{run-and-save-model}(\operatorname{current-date}, \operatorname{hyperparameters}, \operatorname{load-last-model}(\operatorname{last-date}))$
                    \State $\operatorname{last-date} \gets \operatorname{current-date}$
                \EndIf
            \EndFor
        \EndProcedure
    \end{algorithmic}
    \caption{Test Workflow}
    \label{algo:test}
\end{algorithm}

Another important aspect of our testing procedure is how we set up the priors for the weekly simulations (a summary can be seen in Algorithm~\ref{algo:test}).
At any given week, except the first one, posteriors inferred from the previous week were used as starting points for the current models.
This practice of updating the priors with previous estimations is known as Sequential Bayesian Updating, or simply Bayesian Updating \cite{Balayla2022}, and has been used for similar purposes in related literature \cite{Daza-Torres2021}, as well as in other academic fields \cite{Petzschner2011, Bin2019, DeBacco2021-vimure}.
This strategy allowed the Bayesian inference optimisation to converge much faster in our experiments, so we were able to run fewer iterations of the algorithm than if we had to build the model from the ground up every week.
The number of iterations (see Table \ref{tab:hyperparameters}) was chosen after a preliminary test phase where we analysed the trade-off between reliability and execution time.
While this sequential nature of the experiments means that we only report a single run of the model for each week, we still believe the sample size produced is more than enough to allow us to identify which models are better.

\begin{table}[!htb]
\centering
\begin{tabular}{lll}
\toprule
Hyperparameter
    & First model
    & Weekly model \\
\midrule
Warmup Iterations
    & 2400
    & 600 \\

Sampling Iterations
    & 600
    & 600 \\

Number of Chains
    & 4
    & 4 \\

Google Mobility Moving Average Window Size
    & 7
    & 7 \\

Maximum Tree Depth
    & 8
    & 8 \\

Priors
    & As per the baseline models\cite{Flaxman2020}
    & Loaded from previous week \\
\bottomrule
\vspace{0.05em}
\end{tabular}
\caption{\label{tab:hyperparameters}List of hyperparameters used in the tests, with the differences shown between the version used in the first week of modelling vs all the other tests.}
\end{table}

\bibliographystyle{unsrtnat}  
\bibliography{references}

\section*{Acknowledgements}

The research presented in this paper was conceived in the context of the Intersectorial Data Intelligence Center for COVID-19 (NIIDC), a group of volunteers that ran from March until June 2020 from academia, private companies and public local state departments dedicated to discussing data analytic strategies with potential to inform decision-makers in the southern Brazilian state of Santa Catarina (SC) \cite{MPSCNoticia, TJSCNoticia}. 
Special thanks to Dr. Ana Luiza Curi Hallal for the valuable exchanges and discussions about epidemic modelling in the early days of this group.
We thank CIASC (SC) for providing us the credentials to download data from Platform BoaVista.
Thanks, Bang Wong, for providing a colour scheme inclusive for colour-blind people \cite{wong2011points}.
Data Science Brigade acknowledges funding from ICASA (SC). 
P.H.C.A. and L.C.L. acknowledge that this study was financed in part by CNPq (Brazilian Research Council) and Coordenação de Aperfeiçoamento de Pessoal de Nível Superior (CAPES), Finance Code 001.
P.H.C.A. acknowledges that during his stay at KCL and A*STAR he's partly funded by King's College London and by the A*STAR Research Attachment Programme (ARAP).

\section*{Author contributions statement}

P.H.C.A and J.C.S. conceived the experiment(s), P.H.C.A  programmed and conducted the experiment(s), P.H.C.A and J.C.S. analysed the results. All authors reviewed the manuscript. 

\section*{Code availability}

The code for these models can be found on the Github repository: \url{https://github.com/Data-Science-Brigade/modelo-epidemiologico-sc/}.

\clearpage

\section*{Supplementary Information}

\setcounter{figure}{0}
\setcounter{table}{0}
\renewcommand{\thefigure}{S\arabic{figure}}
\renewcommand{\thetable}{S\arabic{table}}

\begin{table}[!ht]
	\centering
	\begin{tabular}{lrrrr}
		\toprule
		\multirow{2}{*}{Model}
		& \multicolumn{2}{c}{RMSE7}
		& \multicolumn{2}{c}{RMSE30} \\
		& pred & conf & pred & conf \\
		\midrule
		
		base
            & 2.55
            & 2.46
            & 5.79
            & 3.57 \\
            
        base-ron
            & 2.29
            & 2.55
            & 2.92
            & 3.06 \\
            
        base-rnn
            & 2.48
            & 2.55
            & 3.14
            & 3.10 \\
            
        base-ror
            & \textbf{2.28}
            & \textbf{2.20}
            & 3.05
            & \textbf{2.82} \\
            
        base-rnr
            & 2.31
            & 2.43
            & \textbf{3.02}
            & 3.01 \\
		\midrule
		
		\multirow{2}{*}{Model}
		& \multicolumn{2}{c}{MAE7}
		& \multicolumn{2}{c}{MAE30} \\
		& pred & conf & pred & conf \\
		\midrule
		
		base
            & 2.19 ($\pm$1.05)
            & 2.09 ($\pm$1.05)
            & 4.77 ($\pm$2.71)
            & 2.96 ($\pm$1.60) \\
            
        base-ron
            & \textbf{1.96} ($\pm$0.95)
            & 2.21 ($\pm$1.01)
            & \textbf{2.40} ($\pm$1.32)
            & 2.51 ($\pm$1.38) \\
            
        base-rnn
            & 2.13 ($\pm$0.98)
            & 2.18 ($\pm$1.06)
            & 2.56 ($\pm$1.45)
            & 2.54 ($\pm$1.41) \\
            
        base-ror
            & \textbf{1.97} ($\pm$0.92)
            & \textbf{1.86} ($\pm$0.96)
            & 2.52 ($\pm$1.36)
            & \textbf{2.27} ($\pm$1.32) \\
            
        base-rnr
            & \textbf{1.97} ($\pm$0.96)
            & 2.07 ($\pm$1.01)
            & 2.48 ($\pm$1.37)
            & 2.45 ($\pm$1.38) \\
		\bottomrule
	\end{tabular}
	\caption{This table shows the error values (RMSE and $avg (\pm std)$ MAE) for the predicted value and their confidence interval counterparts in a 7 and 30-day forecasting window.}
	\label{tab:results}
\end{table}

\begin{table}[!ht]
    \centering
    \begin{tabular}{llrrrr}
        \toprule
        \multirow{2}{*}{Region} & \multirow{2}{*}{Model} & \multicolumn{2}{c}{RMSE7} & \multicolumn{2}{c}{RMSE30} \\
        & & pred & conf & pred & conf \\
        \midrule
        \multirow{5}{*}{Alto Vale do Itajaí}
            & base
                & 2.44
                & 2.09
                & 7.17
                & 3.06 \\
            & base-ron
                & 1.94
                & 2.32
                & 2.39
                & 2.81 \\
            & base-rnn
                & 2.11
                & 2.39
                & 2.69
                & 2.87 \\
            & base-ror
                & 1.91
                & 1.99
                & 2.32
                & 2.50 \\
            & base-rnr
                & 2.13
                & 2.28
                & 2.72
                & 2.79 \\

        \midrule
        \multirow{5}{*}{Foz do Rio Itajaí}
            & base
                & 2.40
                & 2.44
                & 4.61
                & 3.47 \\
            & base-ron
                & 2.14
                & 2.36
                & 2.65
                & 2.79 \\
            & base-rnn
                & 2.21
                & 2.43
                & 3.03
                & 2.96 \\
            & base-ror
                & 2.17
                & 2.12
                & 2.74
                & 2.59 \\
            & base-rnr
                & 2.17
                & 2.32
                & 2.81
                & 2.87 \\
        
        \midrule
        \multirow{5}{*}{Grande Florianopolis}
            & base
                & 2.91
                & 2.75
                & 8.94
                & 4.31 \\
            & base-ron
                & 2.65
                & 2.77
                & 3.43
                & 3.30 \\
            & base-rnn
                & 2.50
                & 2.65
                & 3.34
                & 3.10 \\
            & base-ror
                & 2.67
                & 2.41
                & 3.72
                & 3.10 \\
            & base-rnr
                & 2.49
                & 2.62
                & 3.44
                & 3.06 \\
            
        \midrule
        \multirow{5}{*}{Grande Oeste}
            & base
                & 1.52
                & 1.60
                & 2.96
                & 2.35 \\
            & base-ron
                & 1.64
                & 1.62
                & 2.22
                & 2.26 \\
            & base-rnn
                & 2.93
                & 1.60
                & 3.00
                & 2.29 \\
            & base-ror
                & 1.52
                & 1.46
                & 2.23
                & 2.09 \\
            & base-rnr
                & 1.60
                & 1.52
                & 2.19
                & 2.15 \\
                
        \midrule
        \multirow{5}{*}{Meio Oeste e Serra Catarinense}
            & base
                & 2.10
                & 2.02
                & 3.60
                & 2.75 \\
            & base-ron
                & 1.90
                & 2.23
                & 2.22
                & 2.56 \\
            & base-rnn
                & 1.93
                & 2.32
                & 2.29
                & 2.71 \\
            & base-ror
                & 1.83
                & 1.86
                & 2.27
                & 2.22 \\
            & base-rnr
                & 2.01
                & 2.15
                & 2.35
                & 2.54 \\
        
        \midrule
        \multirow{5}{*}{Planalto Norte e Nordeste}
            & base
                & 2.80
                & 3.14
                & 4.59
                & 4.17 \\
            & base-ron
                & 2.59
                & 2.98
                & 3.37
                & 3.77 \\
            & base-rnn
                & 2.67
                & 3.10
                & 3.69
                & 3.95 \\
            & base-ror
                & 2.58
                & 2.51
                & 3.42
                & 3.34 \\
            & base-rnr
                & 2.63
                & 2.90
                & 3.52
                & 3.77 \\
        
        \midrule
        \multirow{5}{*}{Sul}
            & base
                & 3.66
                & 3.21
                & 8.66
                & 4.86 \\
            & base-ron
                & 3.17
                & 3.58
                & 4.13
                & 3.91 \\
            & base-rnn
                & 3.01
                & 3.37
                & 3.96
                & 3.85 \\
            & base-ror
                & 3.28
                & 3.08
                & 4.62
                & 3.91 \\
            & base-rnr
                & 3.11
                & 3.19
                & 4.10
                & 3.86 \\
        \bottomrule
    \end{tabular}
    \caption{This table shows the error values (RMSE) for the predicted value and their confidence interval counterparts in a 7 and 30-day forecasting window, aggregated over each region the model was ran for.}
    \label{tab:results-regions}
\end{table}

\begin{figure}[!ht]
\centering
    \begin{subfigure}{\textwidth}
        \includegraphics[width=\linewidth]{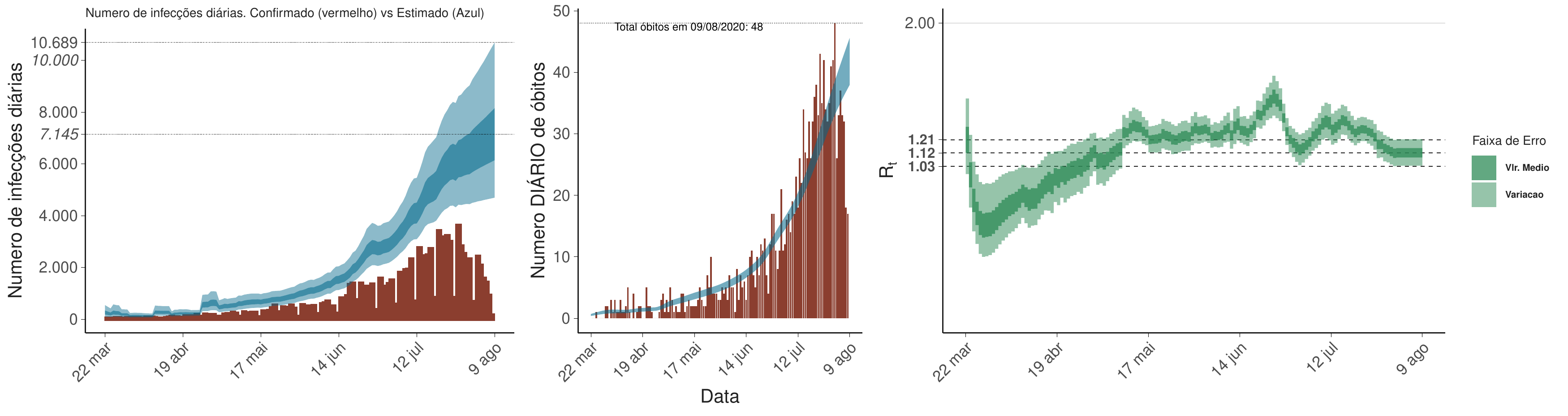}
        \caption{Diagnostic plots for \texttt{base} model for 10/08/2020}
        \label{fig:base-model-abc-state-2020-08-10}
    \end{subfigure}
    \begin{subfigure}{\textwidth}
        \includegraphics[width=\linewidth]{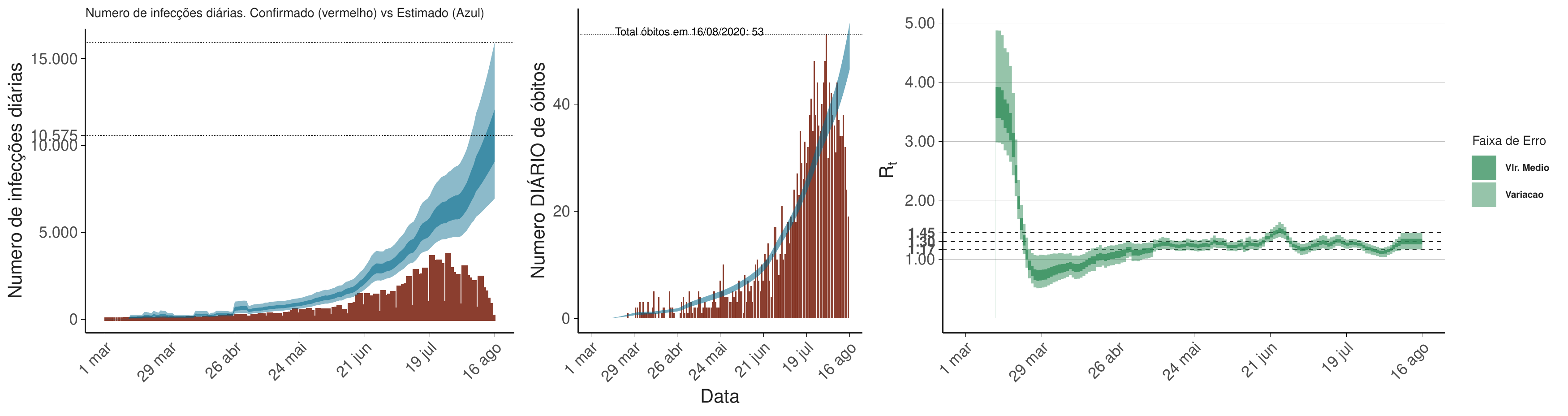}
        \caption{Diagnostic plots for \texttt{base} model for 17/08/2020}
        \label{fig:base-model-abc-state-2020-08-17}
    \end{subfigure}
    \begin{subfigure}{\textwidth}
        \includegraphics[width=\linewidth]{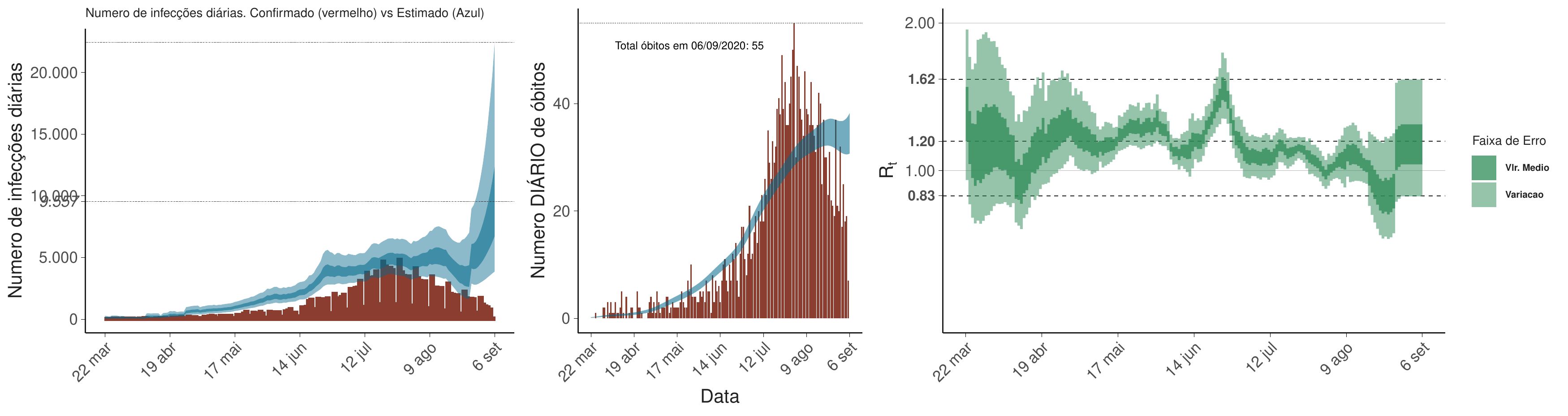}
        \caption{Diagnostic plots for \texttt{base} model for 07/09/2020}
        \label{fig:base-model-abc-state-2020-09-07}
    \end{subfigure}
    \caption{Diagnostics plots for simulations with \texttt{base} model for three selected dates for the state of Santa Catarina. From left to right, the plots indicate: i) number of new reported infections per day (red) and the probable actual number of infections as estimated by the algorithm (blue) ii) number of confirmed deaths per day (red) and the probable actual number of deaths as fitted by the algorithm (blue) iii) the effective reproduction number $R_t$ for the state over the period up until the snapshot date.}
\end{figure}

\begin{figure}[!ht]
\centering
    \begin{subfigure}{\textwidth}
        \includegraphics[width=\linewidth]{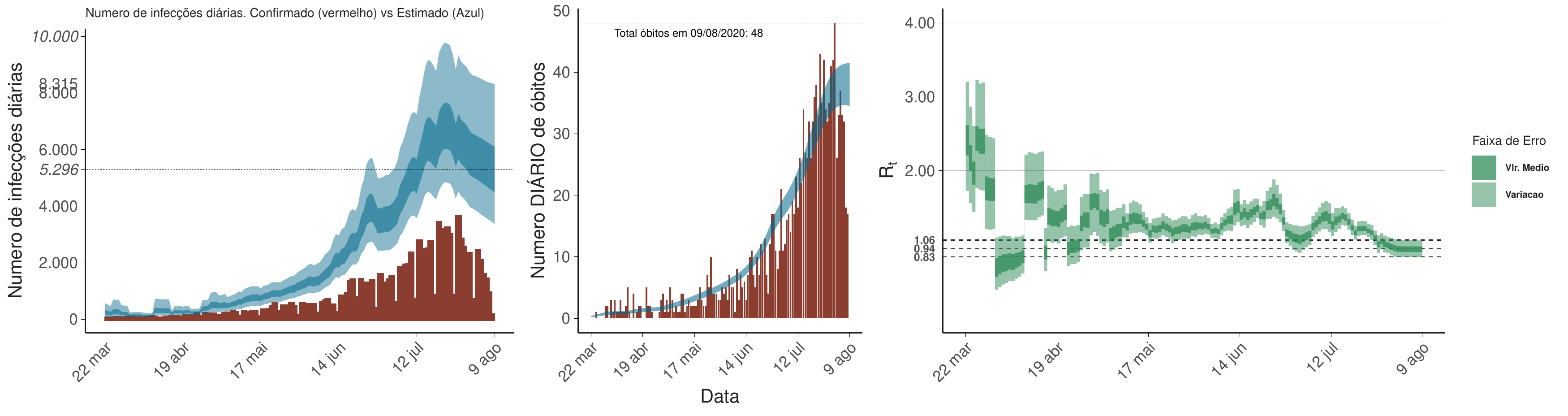}
        \caption{Diagnostic plots for \texttt{base-ron} model for 10/08/2020}
        \label{fig:base-reported-model-abc-state-2020-08-10}
    \end{subfigure}
    \begin{subfigure}{\textwidth}
        \includegraphics[width=\linewidth]{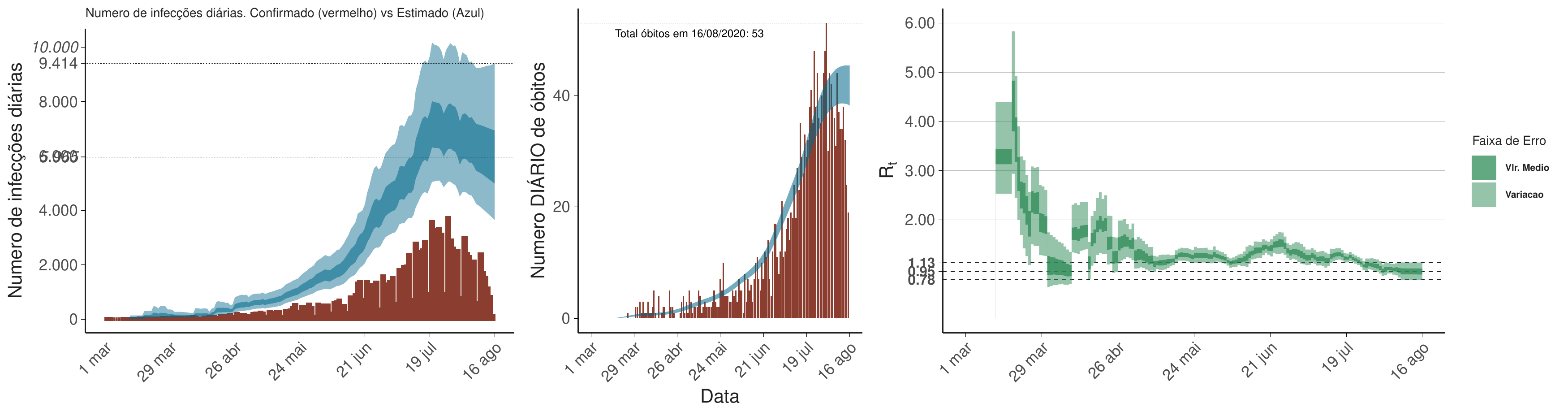}
        \caption{Diagnostic plots for \texttt{base-ron} model for 17/08/2020}
        \label{fig:base-reported-model-abc-state-2020-08-17}
    \end{subfigure}
    \begin{subfigure}{\textwidth}
        \includegraphics[width=\linewidth]{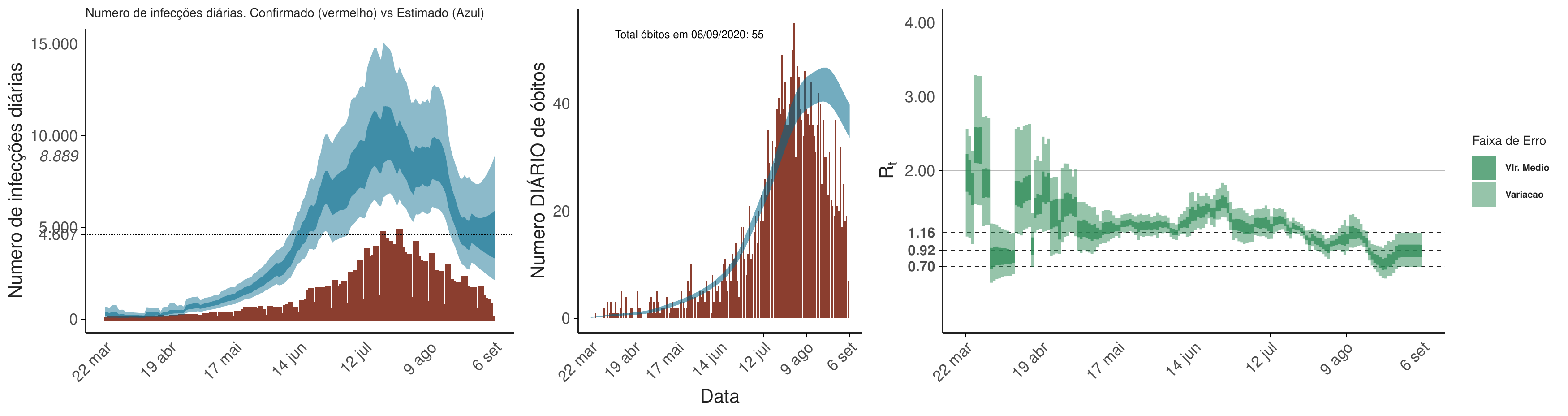}
        \caption{Diagnostic plots for \texttt{base-ron} model for 07/09/2020}
        \label{fig:base-reported-model-abc-state-2020-09-07}
    \end{subfigure}
    \caption{Diagnostics plots for simulations with \texttt{base-ron} model for three selected dates for the state of Santa Catarina. From left to right, the plots indicate: i) number of new reported infections per day (red) and the probable actual number of infections as estimated by the algorithm (blue) ii) number of confirmed deaths per day (red) and the probable actual number of deaths as fitted by the algorithm (blue) iii) the effective reproduction number $R_t$ for the state over the period up until the snapshot date.}
\end{figure}

\begin{figure}[!htbp]
    \centering
    \includegraphics[width=0.8\linewidth]{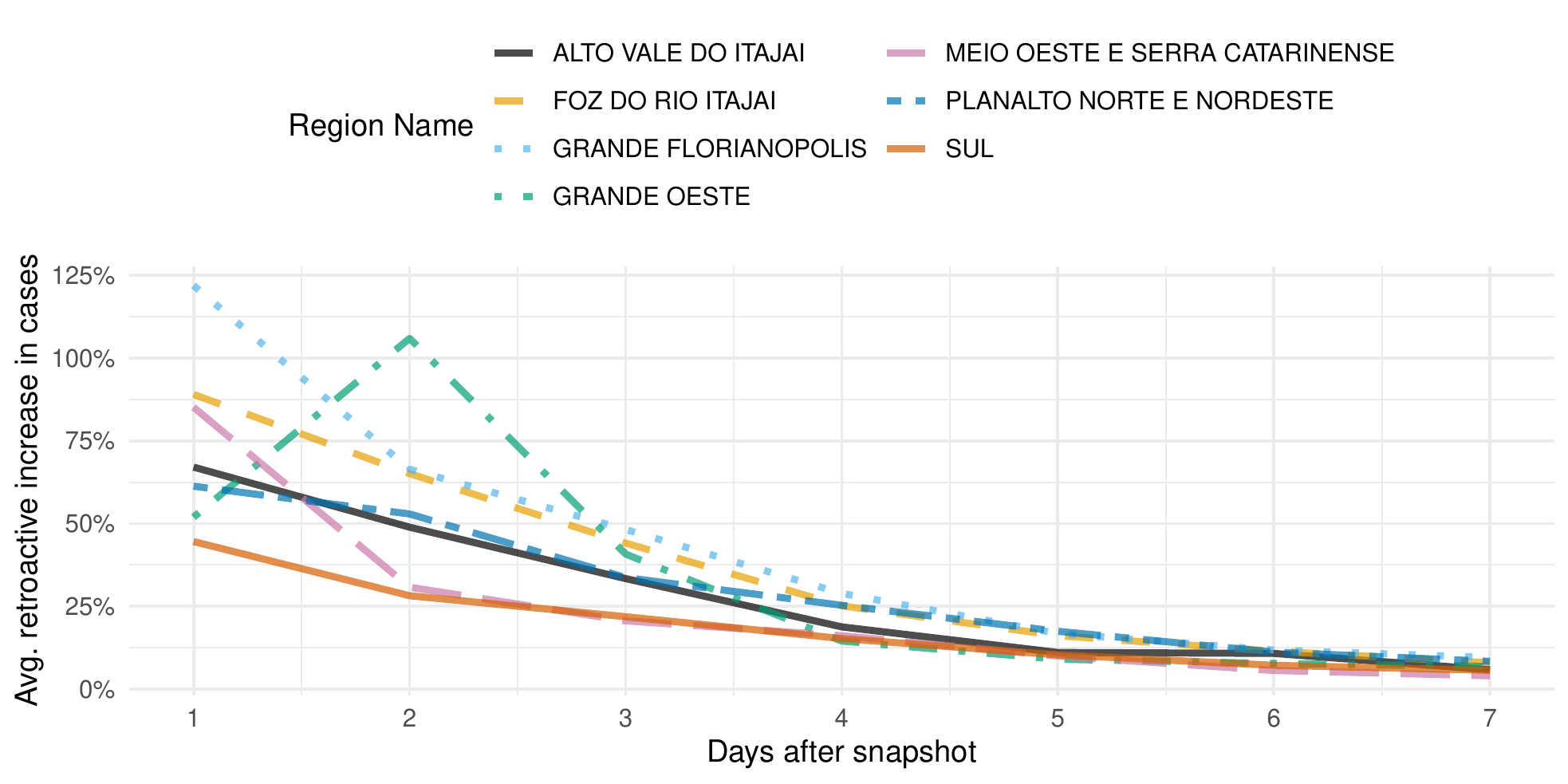}
    \caption{Distribution of retroactively-added cases on the last date. For example, if for a region there are 100\% average retroactive increase in cases 2 days after the snapshot, it means that, on the date two days earlier than the snapshot's date, the number of cases should be doubled to represent the true value.}
    \label{fig:retroactive-cases}
\end{figure}

\begin{figure}[!htbp]
    \centering
    \includegraphics[width=0.8\linewidth]{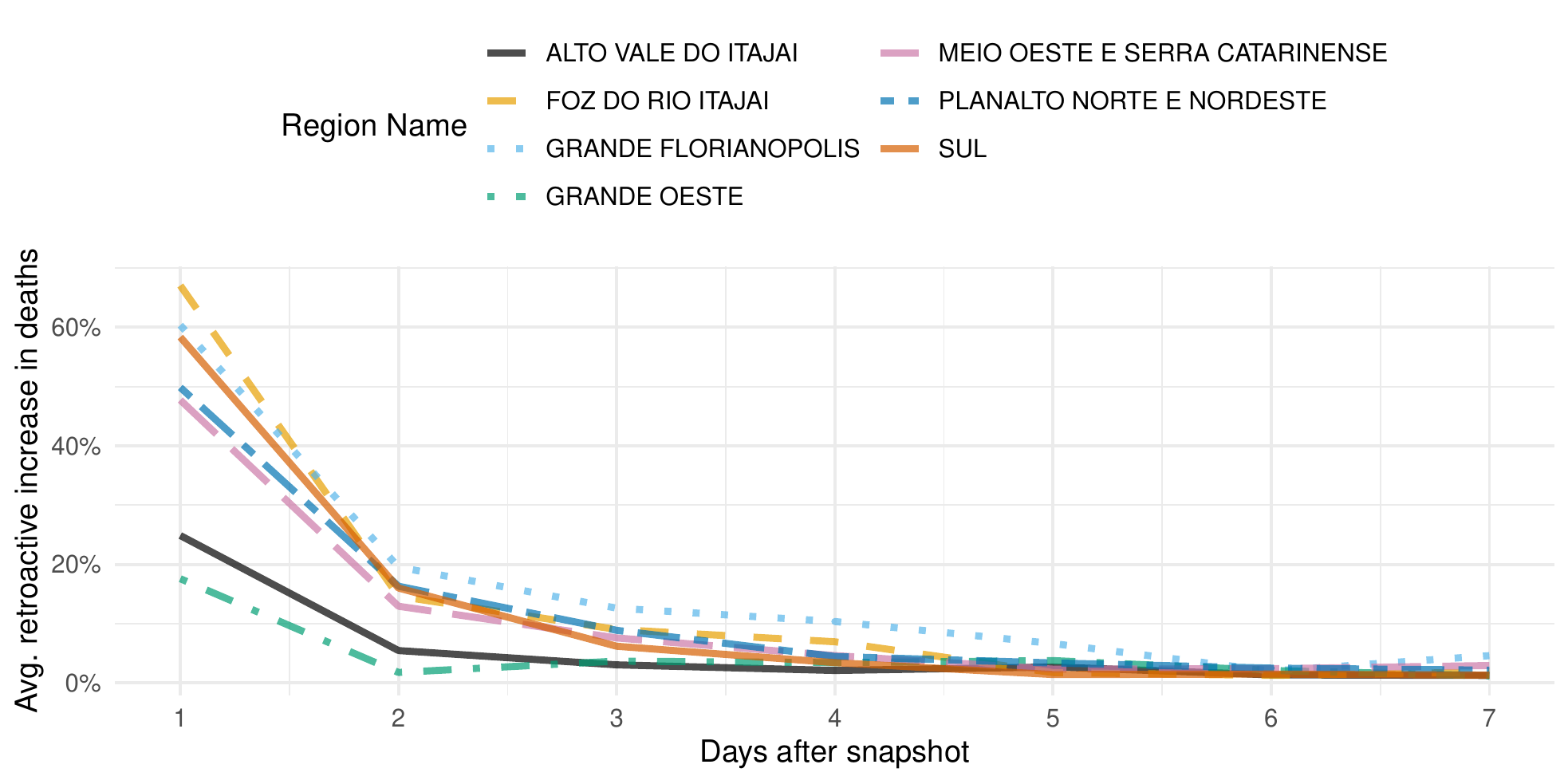}
    \caption{Distribution of retroactively-added deaths on the last date. For example, if for a region there are 100\% average retroactive increase in deaths 2 days after the snapshot, it means that, on the date two days earlier than the snapshot's date, the number of deaths should be doubled to represent the true value.}
    \label{fig:retroactive-deaths}
\end{figure}

\begin{figure}[!htbp]
\centering
\includegraphics[width=0.7\textwidth]{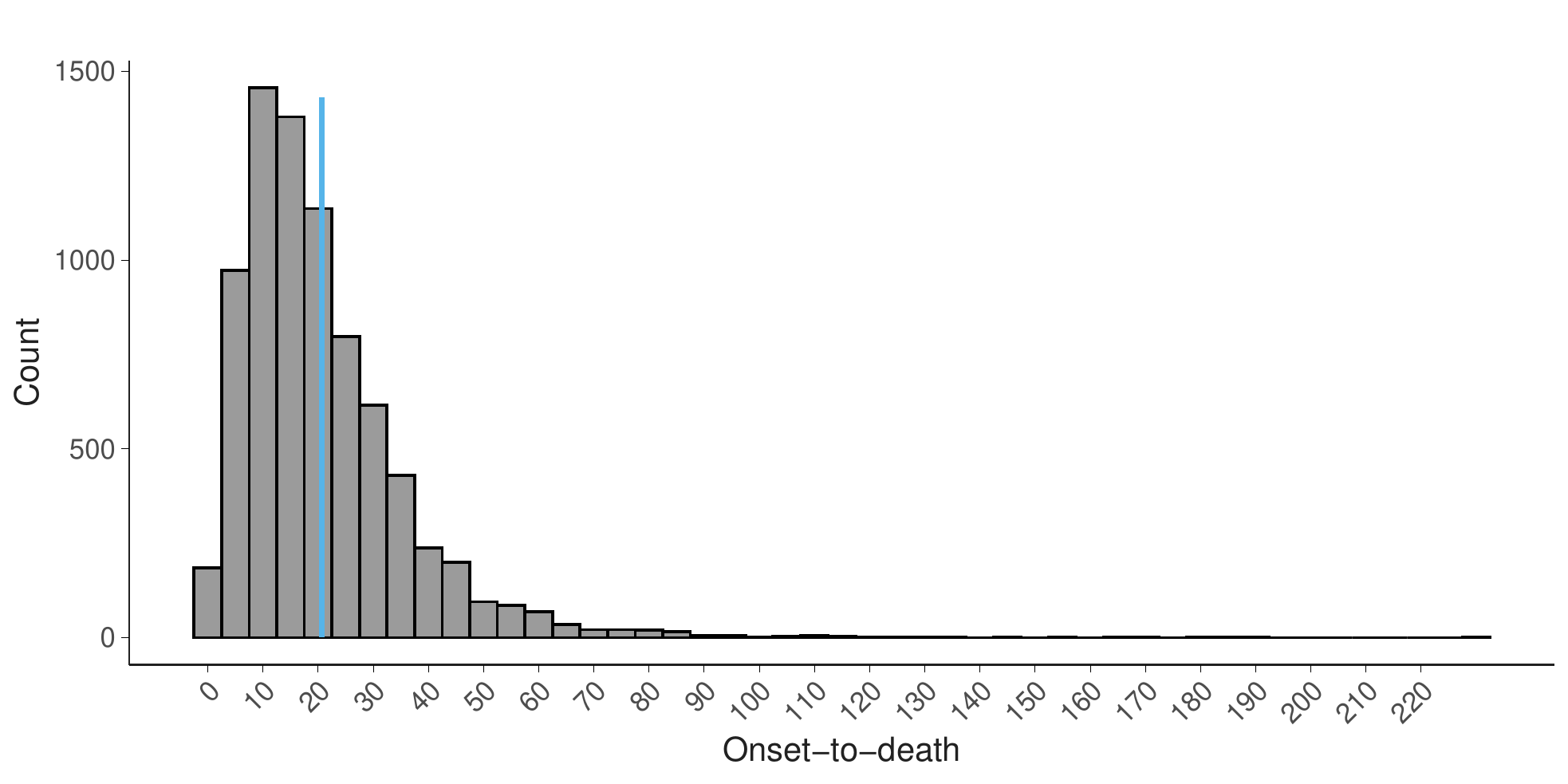}
\caption{Distribution in days relative to the date of onset of first reported symptoms until date of death across Santa Catarina state in Brazil, according to data provided by Plataform Boa Vista \cite{CIASC-SC2021}.}
\label{fig:onset-to-death}
\end{figure}

\begin{figure}[!htbp]
\centering
\includegraphics[width=\textwidth]{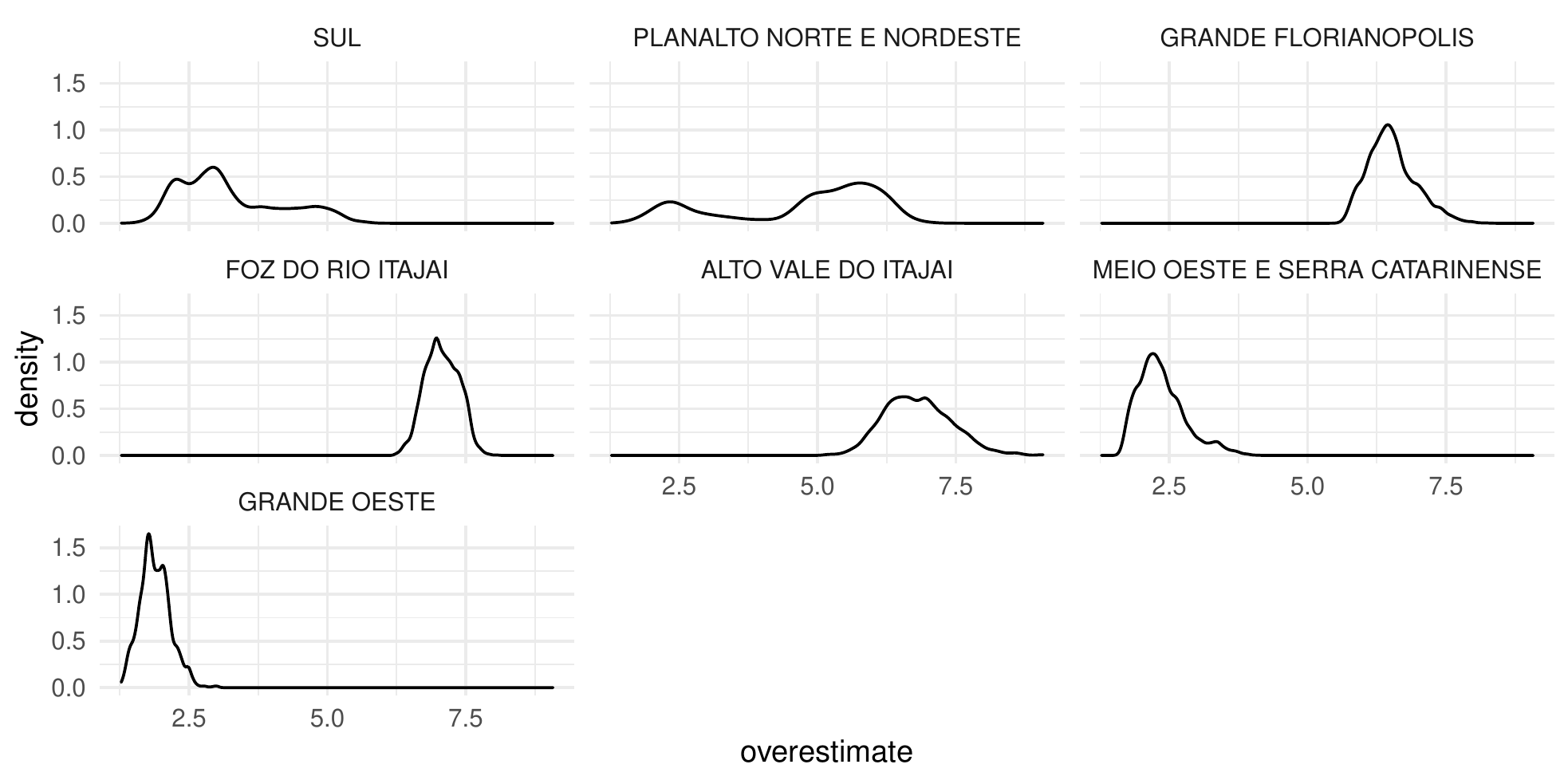}
\caption{Distribution of the overestimate parameter in the last 600 iterations of all chains for the last \texttt{base-ron} model. The multimodality present here is due to the fact that we ran 4 independent chains, which generated different distributions that, when combined, produce a multimodal distribution.}
\label{fig:overestimate}
\end{figure}


\end{document}